\documentclass[aps,prb,amsfonts,amssymb,twocolumn]{revtex4}

\usepackage{graphicx}

\begin{document}

\title{Mean-field glass transition in a model liquid}

\author{V.S.\ Dotsenko$^{\, a,b}$ and G.\ Blatter$^{\, c}$}

\affiliation{$^a$LPTL, Universit\'e Paris VI, 75252 Paris, France} 
\affiliation{$^b$L.D.\ Landau Institute for Theoretical Physics, 
   117940 Moscow, Russia}
\affiliation{$^c$Theoretische Physik, ETH-H\"onggerberg, CH-8093 Z\"urich, 
   Switzerland}

\date{\today}

\begin{abstract}
We investigate the liquid-glass phase transition in a system of
point-like particles interacting via a finite-range attractive
potential in $D$-dimensional space. The phase transition is
driven by an `entropy crisis' where the available phase space 
volume collapses dramatically at the transition. 
We describe the general strategy underlying the first-principles 
replica calculation for this type of transition; its application 
to our model system then allows for an analytic description 
of the liquid-glass phase transition within a mean-field 
approximation, provided the parameters are chosen suitably. 
We find a transition exhibiting all the features associated
with an `entropy crisis', including the characteristic finite
jump of the order parameter at the transition while the free 
energy and its first derivative remain continuous.
\end{abstract}

\maketitle

\section{Introduction}\label{sec:intro}

Notable progress in understanding fundamental aspects of
structural glasses and their freezing transition
has been made \cite{glass}; significant advancements
originate from the dynamic formulation of the glass
transition using mode coupling theory \cite{modecoupling},
while the statistical mechanics approach
draws extensively from the analogy between
the well studied spin-glass \cite{mpv} and
the glassy solid \cite{K}. While the spin glass is
characterized by random frozen orientations of spins
due to the presence of quenched disorder, the
structural glass is characterized by random
frozen space positions of the particles but does not
rely on the presence of quenched disorder. A central idea
put forward in this context is the concept of the
`entropy crisis' \cite{GibbsDiMarzio} driving the transition
into the glassy state through a collapse of the
system's phase space. Transitions of this type have
shown up in various disordered systems \cite{rem,1rsb}
and in the `discontinuous' spin-glasses containing 
no quenched disorder \cite{repl} (see also Ref.\ 
\cite{glass1} for similar systems with other then
spin degrees of freedom).
In this paper, we apply a heuristic framework\cite{MP}
based on an `entropy crisis' scenario to describe
the liquid-glass transition and the low-temperature
thermodynamics of the glassy state in a system
of interacting particles in $D$-dimensional space.

\begin{figure}[ht]
\includegraphics[scale=0.4]{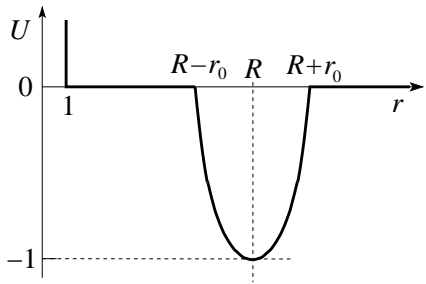}
\caption[]{Interaction potential between particles used
   in our model glass former. The attractive interaction
   (normalized to unity) is limited to a shell of width
   $2 r_0$ at a radius $R \gg r_0\gg 1$, where we have
   set the particle radius to unity. The repulsive core
   prevents the collapse of the system into two clusters.}
  \label{fig:ur}
\end{figure}

Progress in our understanding of the liquid-glass phase 
transition and the physics of the low-temperature glass 
state is made along two avenues: {\it i)} experimental 
and numerical studies provide new details on specific 
materials and on model systems but have little impact 
on our general understanding of the glass phenomenon. 
On the other hand, {\it ii)} conceptual studies push 
our general understanding but unfortunately provide 
little predictive power when it comes to the description 
of realistic systems. In choosing a suitable model 
system, we then have to compromise between a realistic 
description of the glass former and one allowing us 
to make analytical progress, e.g., within a mean-field 
approach. The first model describing a liquid-glass 
phase transition and allowing for a mean-field type 
analysis was proposed by Kirkpatrick and Thirumalai 
\cite{KT}. Formulated within a density functional 
theory, it provided a consistent static and dynamic 
description of the structural glass transition. New 
insights into the nature of the glass transition
based on studies of coupled replicated glassy systems
\cite{Monasson,FranzParisiCardenas} led to the 
formulation of a first principles computational 
scheme providing a description of the equilibrium 
thermodynamics of glasses \cite{MP}. This scheme 
has been successfully applied, combining analytical 
and numerical techniques, to the soft-spheres model 
in three dimensions \cite{MP} and to the Lennard-Jones 
binary mixture \cite{CPV}. Recently, a model with 
point-like particles interacting via a spatially 
oscillating infinite-range potential has been analyzed
within this framework \cite{mfg1}. Despite its 
unrealistic structure, this model turned out to
provide a good testbed for the replica based approach:
the physics of the liquid-glass phase transition 
revealed in this model turned out fully consistent 
with the heuristic ideas of the `entropy crisis' 
scenario. In the present paper, we consider a much 
more realistic model of a structural glass: adopting 
again the scheme suggested in Ref.\ \onlinecite{MP},
we study the low-density limit of a system of 
particles in $D$-dimensional space interacting via a 
finite-range attractive potential of depth $U_0$, 
confined to a shell of radius $R$ and thickness $2 r_0$, 
see Fig.\ \ref{fig:ur}.  It turns out, that an 
appropriate choice of the interaction parameters 
($r_0 \ll R$) allows us to adopt a mean-field 
approximation and proceed with an analytic calculation 
of the free energy and a properly defined order 
parameter to the very end. We then arrive at a 
complete analytic description of the liquid-glass 
phase transition occurring in this model glass
former and demonstrate that it is again consistent 
with the `entropy crisis' scenario.

In spite of this success, we have to admit ignoring various
important issues related to the structural glass transition: 
First, we explicitly avoid the discussion of the relevance 
of our equilibrium statistical mechanical approach for the 
apparently non-equilibrium type liquid-glass phase transition. 
This question is both general and deep and we are unable to 
present an answer at the present stage. Rather, we hope that, 
like in the case of spin-glasses (another example of matter 
residing in a non-equilibrium state), at least some of the 
physical phenomena and observables are well defined and make 
proper physical sense even when computed in terms of an 
equilibrium approach.  Second, we also prefer not to start 
a lengthy (and useless) discussion of the question, what 
to do with the {\it crystal} state. Such a crystalline state 
is certainly present in our model and, since its energy is 
evidently lower than the typical energy of glassy configurations, 
it is the true ground state of the system. In fact, this 
is the typical situation for most models of glasses and 
the usual algorithm of treating the presence of the crystal 
state (which we also follow in this paper) is simple: 
it has to be ignored. On a qualitative level, the reason 
for such a pragmatic approach is very simple:
it is well known that frozen glassy states do exist at low
temperatures regardless of the presence of the crystal ground
state. Moreover, in many circumstances such states turn out to
be quite stable during reasonable observation times; we then can
safely assume that the crystal state is located far away
from the relevant glassy states in configurational space
and furthermore that the crystal and glassy states are separated
by large energy barriers.  This is a quite standard
situation in statistical mechanics: except for some rare cases
admitting an exact solution, one usually studies only a limited
{\it ad hoc} chosen part of the phase space depending on the
object under study. In the case of the structural glass,
one then expects two scenaria: either the basic assumption
(that the existence of the crystal can be ignored) turns out
to be reasonable and the crystal configurations never show up
in the calculations. Or, this assumption is wrong and then
one inevitably faces some kind of instabilities and
divergences. As for the model and method considered in the
present paper, the calculations demonstrate that the crystal
state indeed does not interfere with the glassy configurations.

The paper is structured as follows:
In section II below, we first describe the heuristic 
framework underlying our replica analysis along the 
lines discussed in Ref.\ \onlinecite{MP} and introduce
our specific model. Its free energy is calculated 
within the replica mean-field approach in section III.
The results and conclusions are given in section IV:
We find that the system freezes at the glass temperature 
$T_c = U_0/\ln(R/r_0)$ into an amorphous solid with a 
number of nearest neighbors slightly larger than $D$. 
We determine the order parameter and its jump at $T_c$ 
and present analytical expressions for the free energy 
and entropy of the solid and liquid phases as well as 
for the configurational entropy or complexity.

\section{Structural glass transition}\label{sec:glass}

\subsection{Symmetry breaking in random systems}

Consider a system of $N$ particles in $D$-dimensional space described
by the Hamiltonian
\begin{equation}
   \label{mfg1}
   H[{\bf x}_{i}]
   = \frac{1}{2} \sum_{i,j=1}^{N} U(|{\bf x}_{i} - {\bf x}_{j}|),
\end{equation}
where ${\bf x}_{i}$ denotes the position of the $i$-{\it th} 
particle and $U(|{\bf x}|)$ is the interparticle potential. 
We {\it assume} that at low temperatures the system is 
frozen in a disordered (glassy) state which is characterised 
by random spatial positions of the particles; this can be 
achieved through a rapid quench avoiding the crystallization 
by entropic reasons. The glass state is characterized by 
broken translational and rotational symmetries, but 
unlike the ordered crystal configurations (characterized 
by their specific spatial and
rotational symmetries), it is impossible to identify the 
residual symmetries left in the glassy state. This naturally
resembles the spin-glass problem, where the spins are frozen 
in a random state which cannot be characterized by any 
apparent global symmetry breaking. However, unlike spin-glasses,
here we do not have {\it quenched disorder} installed in the 
initial Hamiltonian. Nevertheless, the ideas borrowed from 
the spin-glass theory and in particular the use of the replica 
technique turn out to be quite fruitful also for 
the description of structural glasses \cite{MP}.

In order to demonstrate the effect of spontaneous symmetry
breaking, e.g., in ordered magnetic systems, one introduces a
conjugate field coupled to the order parameter which is set
to zero at the end (after taking the thermodynamic limit). In
spin-glasses, the same strategy can be applied by introducing several
weakly coupled copies (replicas) of the original system. Similarly,
in order to demonstrate the freezing of a system of interacting
particles (described by (\ref{mfg1})) into a random glass state, we 
introduce two identical copies (with particles at positions 
${\bf x}_{i}$ and ${\bf y}_{i}$, respectively) of the same 
system described the Hamiltonian
\begin{eqnarray}
   \label{mfg2}
   H_2 &=& \frac{1}{2} \sum_{i,j}^{N} U(|{\bf x}_{i} - {\bf x}_{j}|)
   + \frac{1}{2} \sum_{i,j}^{N} U(|{\bf y}_{i} - {\bf y}_{j}|)
   \nonumber\\
   && \qquad\qquad\qquad
   + \epsilon \sum_{i}^{N} W({\bf x}_{i} - {\bf y}_{i}).
\end{eqnarray}
The last term in (\ref{mfg2}) describes a weak attractive potential 
$W$ between the particles at ${\bf x}_i$ and ${\bf y}_i$ of the 
two systems and plays the role of the symmetry breaking 
conjugate field in the ordered system. As usual,
the control parameter $\epsilon$ is set to zero
{\it after} taking the thermodynamic limit and
the system can end up in one of two phases:
{\it i)} the particles of the two systems are
independent (uncorrelated), indicating that they  
do not memorize their spatial positions and hence
the original system is in the high-temperature
liquid phase, or {\it ii)} the positions of the
particles remain correlated, indicating that they
are localized in space and we conclude that the
original system is in the low-temperature solid state.
As in spin glasses, in order to obtain more detailed
information about the phase transition, it is convenient
to introduce $m$ replicas of the original system.  
Also, in the actual calculation there is no need
to introduce a supplementary attractive potential
between the replicas: following standard practice,
it is sufficient to allow for the possibility of
symmetry breaking in order to prove its existence
afterwards.

\subsection{Entropy crisis scenario}

The entropy crisis scenario \cite{GibbsDiMarzio} for 
the glass transition builds on the idea of a phase 
space collapse upon entering the frozen phase; in
its pure form it shows up in the random energy
model of spin glasses which has been solved exactly
\cite{rem}. We first briefly summarize the main 
features of this heuristic framework as applied to 
the problem of structural glasses, following the
original work of M\'ezard and Parisi \cite{MP}.

\begin{figure}[h]
\includegraphics[scale=0.40]{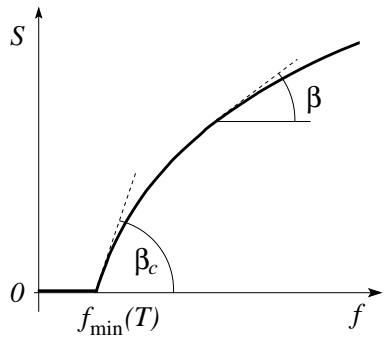}
\caption[]{Qualitative shape of the configurational 
  entropy density assumed in the phenomenological 
  description of a glass transition driven by an 
  entropy crisis. The slope $\beta$ corresponds to 
  the inverse temperature and assumes its maximal 
  value $\beta_c =1/T_c$ at the glass transition 
  temperature $T_c$.}
  \label{fig:sf}
\end{figure}

We assume that the partition function $Z$ can be 
represented in the form
\begin{equation}
   \label{Z}
   Z = \sum_{\alpha} \exp\left(-N f_{\alpha}/T \right),
\end{equation}
where $N f_{\alpha}$ denote the energies of the
thermodynamically relevant local mimima in
configurational space. The number $\Omega(f)$ of
local minima with energy $f$ is assumed to be
exponentially large, $\Omega(f) = \exp[ N S(f,T)]$,
where $S(f,T)$ denotes the configurational entropy
density or complexity. Finally, the function $S(f,T)$
at fixed $T$ shall have the qualitative shape shown in
Fig.\ \ref{fig:sf}, with $S(f,T) = 0$ at $f \leq 
f_{\mathrm{min}}(T)$, and a concave increase for 
$f > f_{\mathrm{min}}(T)$.

The partition function is written in the form
\begin{equation}
   \label{mfg5}
   Z \approx \int_{f > f_{\mathrm{min}}} df \exp\bigl\{-N
   [f - T S(f,T)]/T \bigr\}
\end{equation}
which can be evaluated within a saddle-point approximation
in the thermodynamic limit ($N \to \infty$). The free
energy density $F(T) = -T ({\rm ln}Z)/N$ is given by
\begin{equation}
\label{F>}
F(T)  =   f_{*}(T) - T S[f_{*}(T),T], \qquad T > T_c,
\end{equation}
with $f_{*}(T)$ defined via $\partial_f S|_{f_*(T)}
= 1/T < \beta_c$ and $\beta_c$ the maximal slope of the
function $S(f)$, cf.\ Fig.\ \ref{fig:sf}. At low
temperatures $T < T_c$ the integral is determined by the
lowest energy state alone, $F(T) = f_{\mathrm{min}}(T)$.

A microscopic formalism capturing the above phenomenology
can be set up with the help of replicas \cite{MP}:
Consider $m$ identical (non-coupled) replicas of the
same system where the particle positions remain
correlated among the different replicas; this Ansatz
describes a molecular liquid where each molecule
consists of $m$ particles originating from different
replicas. The partition function of the replicated
system takes the form
\begin{equation}
   \label{mfg9}
   Z_{m} \approx \int_{f > f_{\mathrm{min}}} df
   \exp\bigl\{-N m [f-(T/m) S(f,T)]/T \bigr\}.
\end{equation}
Note that the phase space volume $\Omega(f)$ remains that
of the non-replicated liquid since the molecular structure
essentially preserves the configurational degrees of
freedom, hence the `entropic temperature' is reduced
by a factor $m$. The free energy density $F(m,T) = -(T/mN)
\, {\rm ln}Z_{m}$ of the molecular liquid reads
\begin{equation}
\label{Fml}
   F(m,T) \!= \!\left\{
   \begin{array}{ll}
   \displaystyle{\!\! f_{*}(m,T)-\frac{T}{m} S[f_{*}(m,T),T]},
   & \!T > m T_c,
   \\ \noalign{\vskip 5 pt}
   \!f_{\mathrm{min}}(T), 
   & \!T < m T_c,
   \end{array} \right.
\end{equation}
where $f_{*}(m,T)$ is defined via the saddle-point equation
\begin{equation}
   \label{dS}
   \partial_f S(f,T)\big|_{f_*(m,T)} = m/T < \beta_c.
\end{equation}
In order to study the thermodynamic properties of our
original system (with $m = 1$) in the low temperature
glassy phase at $T < T_c$, we continue analytically
the expression for the free energy density $F(m,T)$
from integer values $m$ to arbitrary continuous
values and analyze its behavior for $m < 1$.
Starting from small $m$ with $0 < m < T/T_{c} \equiv
m_{*}(T)$, the $m$-replica system resides in the
molecular liquid phase with a free energy density
given by (\ref{Fml}). As $m$ approaches $m_{*}(T)$
from below, the free energy density becomes pinned
to the lowest value $f_{\mathrm{min}}(T)$ and the
system freezes into the glassy phase. A further
increase of $m$ beyond $m_{*}(T)$ results in a
constant free energy density $f_{\mathrm{min}}(T)$
and this remains to be the case as $m \to 1$. As
a result, the free energy density (and hence the entire
thermodynamics) of the original glass phase can be
computed from the $m$-fold replicated system residing
in the molecular liquid phase and taking the limit
$F_{m=1}(T) \equiv F[m_{*}(T),T]$. The critical
temperature $T_c$ is reached when $m_{*} (T_c) = 1$.
At temperatures above $T_c$ the system resides in
a liquid phase: replicas are independent, hence 
$m=1$, and the free energy is given by $F_{m=1}(T) 
= F[1,T]$.

\begin{figure}[h]
\includegraphics[scale=0.40]{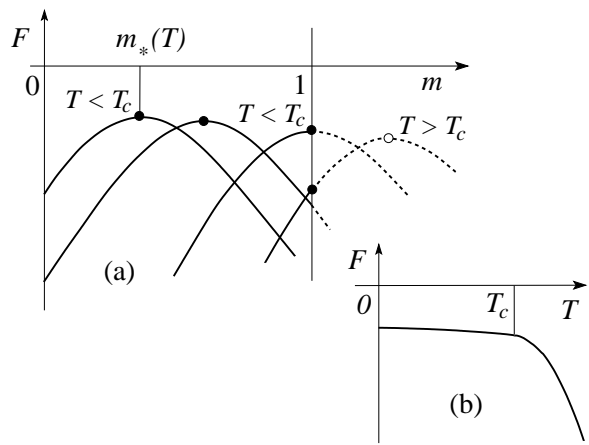}
\caption[]{(a) Sketch of the free energy density $F(m,T)$ for a
  $m$-fold replicated system as a function of the replica
  parameter $m$ for different values of the temperature $T$.
  The smooth decrease of the maximum in $F(m)$ with increasing
  temperature $T$ reflects the (weak) temperature dependence
  of $f_{\mathrm{min}}(T)$.
  (b) Free energy density as a function of $T$ after eliminating
  the replica parameter $m$.}
  \label{fig:fm}
\end{figure}

The crucial step then is the determination of $m_*(T)$.
Assume we have managed to compute the free energy
density $F(m,T)$ of the $m$-replica system for the
molecular liquid phase (note the crucial role of
ergodicity in the liquid allowing for an unrestricted
averaging over phase space; we cannot hope to do such
a calculation for the amorphous solid with its
restricted phase space). Provided our system indeed
follows the heuristics of an entropy crisis, we can
cast this free energy density into the form (\ref{Fml})
with $f_{*} (m,T)$ the solution of the saddle-point
equation (\ref{dS}). The critical replica parameter
$m_{*}(T)$ then is determined by the condition
$f_{*}(m_{*}(T),T) = f_{\mathrm{min}}(T)$ or,
equivalently, $S[f_{*}(m_{*}(T)),T] = 0$.
Fortunately, we do not need to know {\it a priori}
the form of the configurational entropy density $S(f,T)$:
calculating the derivatives of the free energy density
$F(m,T)$ with respect to $m$ at $m = m_{*}(T)$, we
easily find that 
\begin{eqnarray}
   &&\partial_m F(m,T)\big|_{m=m_{*}}\!=
   \big\{(T/m^2) S[f_{*}(m,T),T]
   \nonumber \\
   &&\qquad\quad+\,(\partial_m f_{*})\,\partial_{f_{*}}
   \big[f_{*}-T S(f_{*},T)/m\big]\big\}_{m=m_{*}}\!\!\! = 0
   \nonumber
\end{eqnarray}
and 
\begin{eqnarray}
   \partial^{2}_m F(m,T)\big|_{m=m_{*}}=
   \big[(\partial_m f_{*})/m\big]_{m=m_{*}}\!< 0,
   \nonumber
\end{eqnarray}
where we have made use of the specific shape 
of the function $S(f,T)$, cf.\ Fig.\ \ref{fig:sf}.
We thus conclude that the function $F(m,T)$ exhibits
a {\it maximum} at $m=m_{*}$, cf.\ Fig.\ \ref{fig:fm},
and we can determine the critical replica parameter 
$m_*(T)$ directly from the free energy density $F(m,T)$ 
in the molecular liquid phase without explicit knowledge 
of the entropy density $S(f,T)$.

In summary, we can adopt the following strategy in order
to obtain a proper thermodynamic description of the glass 
transition associated with an entropy crisis: 
First, we have to compute the free energy
density $F(m,T)$ of the $m$-fold replicated system in the
molecular liquid phase and for arbitrary values of the
parameter $m$. Next, this function has to be analytically
continued to arbitrary (non-integer) values of $m$, in
particular, values within the interval $0 < m < 1$. In a
last step, we have to find the maximum of the function
$F(m,T)$ within the interval $[0,1]$. If the maximum
of the replica free energy $F(m,T)$ is realized at
$m_{*}(T) < 1$ (i.e., $T < T_c$, cf.\ Fig. \ref{fig:fm}),
we conclude that the original system (with $m=1$) is in
the glass phase and its free energy density is given
by $F_{\mathrm{glass}} = F[m_{*}(T),T]$. On the other hand,
a maximum in $F(m,T)$ located at a value $m_{*}(T) > 1$
(i.e., $T_c < T$) tells us that the original system resides
in the liquid phase. The true free energy density of the
original system with $m=1$ then is given by the value
$F_{\mathrm{liquid}} = F(m=1,T)$ assumed at the border 
of the interval. Finally, the transition from one 
regime to the other at $m_*(T_c) = 1$ determines the 
transition temperature $T_c$.

\subsection{Implementation of replica calculation}

We consider a system of $N$ identical particles, 
confined within a macroscopic box of volume $V$ 
and described by the Hamiltonian
$H[{\bf x}_{1},{\bf x}_{2},...,{\bf x}_{N}]
\equiv H[{\bf x}]$. The partition function for $m$ 
uncoupled copies of this system reads
\begin{equation}
   \label{mfg18} 
   Z_{m} \!=\! \frac{1}{(N!)^{m}}
   \prod_{i=1}^{N}\prod_{a=1}^{m} 
   \int \! \frac{d^{D} {\bf x}_{i}^{a}}{s^{D}}
   \exp\Bigl[-\!§\beta \sum_{a=1}^{m} H[{\bf x}^{a}]\Bigr],
\end{equation}
where $\beta = 1/T$ denotes the inverse temperature and $s$ 
is the particle size (or the lattice spacing) which we set
to unity hereafter, $s = 1$. Also, we define the density 
$\rho = N/V$, which we keep constant in the thermodynamic 
limit, as well as the mean interparticle distance $L = 
\rho^{-1/D}$. The free energy density is given by the 
expression $F(m,\beta)=-({\rm ln}Z_{m})/m \beta N$.
Following the above strategy, we assume that the spatial 
positions of the particles in different replicas are 
correlated, i.e., the particles arrange in `replica molecules';
technically this is implemented via a transformation to
center of mass (${\bf x}_{i}$) and relative coordinates
(${\bf u}_{i}^{a}$), ${\bf x}_{i}^{a} = {\bf x}_{i} + 
{\bf u}_{i}^{a}$, and assuming that the displacements 
${\bf u}_{i}^{a}$ remain bounded, $|{\bf u}_{i}^{a}| \ll 
L$. In addition, the coordinates have to satisfy the constraint
\begin{equation}
   \label{mfg21} 
   \sum_{a=1}^{m} {\bf u}_{i}^{a} = 0.
\end{equation}
Finally, the assumption that this replica system resides in the 
molecular liquid state implies that the molecular positions 
${\bf x}_{i}$ take on arbitrary values within the entire 
system volume. Hence, the replicated partition function can be
written in the form
\begin{eqnarray}
   \label{mfg22} 
   Z_{m} &=& \frac{m^{DN}}{N!}
   \prod_{i=1}^{N}
   \prod_{a=1}^{m} \int d^{D} {\bf u}_{i}^{a}
   \, \delta\Bigl(\sum_{b=1}^{m} {\bf u}_{i}^{b}\Bigr)
   \nonumber \\ && \!\!\!
   \times \prod_{i=1}^{N}\int d^{D} {\bf x}_{i}
   \exp\Bigl[-\beta \sum_{a=1}^{m} H[{\bf x} 
   + {\bf u}^{a} ]\Bigr].
\end{eqnarray}
Here, the additional factor $m^{DN}$ is due to the change of 
variables ${\bf x}_{i}^{a} = {\bf x}_{i} + {\bf u}_{i}^{a}$
with the constraint (\ref{mfg21}). 

Our problem now has assumed a form which is similar to the replica 
representation of usual glass problems with quenched disorder. The 
molecular coordinates ${\bf x}_{i}$ play the role of the disorder 
parameters, while the displacements ${\bf u}_{i}^{a}$ represent 
the dynamical variables. Our task then is to average over the 
disorder parameters $\{{\bf x}_{i}\}$ in order to arrive at 
the partition function $Z_{m}$ expressed through a new effective 
{\it replica Hamiltonian} $H_{m}[{\bf u}^{a}]$,
\begin{equation}
   \label{mfg23} 
   Z_{m} = \prod_{i=1}^{N}\prod_{a=1}^{m} \int d^{D} 
   {\bf u}_{i}^{a}\, \delta\Bigl(\sum_{b=1}^{m}   
   {\bf u}_{i}^{b}\Bigr) 
   \exp\left(-\beta H_{m}[{\bf u}^{a}]\right),
\end{equation}
where the replica variables ${\bf u}_i^a$ usually become
coupled.  The final integration over the dynamical 
variables ${\bf u}_i^a$ then will provide us with 
the free energy density $F(m,\beta)$. Note that in order 
to assure the consistency of the calculation, we have 
to compute the mean squared displacement amplitude $\langle 
{\bf u}^{2}\rangle^{1/2}$ and verify that it is indeed
smaller than the average distance between the particles,
$\langle {\bf u}^{2}\rangle^{1/2} \ll L$. The above program then
is carried out for the specific interaction introduced
in the next section.

\subsection{Model interaction}\label{sec:model}

We define the inter-particle interaction in the Hamiltonian 
(\ref{mfg1}) in terms of the spherically symmetric potential 
$U(|{\bf x}|)$ (cf.\ Fig.\ \ref{fig:ur}, we define 
$R_{\scriptscriptstyle \pm} = R \pm r_0$) 
\begin{equation}
   \label{mfg24} 
   U(|{\bf x}|) \! =\!\!
   \left\{\begin{array}{ll} 
        \displaystyle{\!\Bigl[\frac{(|{\bf x}| - R)^2}{r_0^2} - 1\Bigr]},
        &R_{\scriptscriptstyle -} < |{\bf x}| < R_{\scriptscriptstyle +}, 
        \\ \noalign{\vskip 5 pt}
        0,  
        &\!\!|{\bf x}| \not\in \left[R_{\scriptscriptstyle -},
        R_{\scriptscriptstyle +}\right].
        \\ \end{array} \right.
\end{equation}
We require the thickness $2 r_{0}$ of the spherical 
attractive shell to be small compared to the radius 
$R$, $r_{0} \ll R$. At the same time, we choose a
shell thickness $r_{0}$ large compared to the particle 
size, which we choose equal to unity for simplicity, 
${r_0} \gg 1$; this allows us to perform all the 
calculations within the continuum limit. More specifically,
we will treat the particles as point-like objects when 
integrating over positions. On the other hand, a proper
hard core repulsion is required in order to inhibit 
the trivial collapse of the system into a pair of 
clusters separated by $R$ with $N/2$ particles each.

For later convenience we introduce the volume integral
\begin{eqnarray}
   \label{mfg25} 
   \int d^D {\bf x} \, U(|{\bf x}|) 
   &=& \int_{R_{\scriptscriptstyle -}\leq |{\bf x}|
   \leq R_{\scriptscriptstyle +}} 
   \!\!\!\!\!\!\!\!\!\!\!\!\!\!\!\!\!\!\!\!
   d^D{\bf x} \quad\  \left[\frac{(|{\bf x}|-R)^2}
   {r_0^2} - 1\right] \nonumber \\
   \noalign{\vskip 3pt}
   &\simeq& - V_{0} \sim R^{D-1} r_{0}
\end{eqnarray}
with $V_{0} = 2 r_{0} S_{D}$ the volume of a spherical shell 
with radius $R$ and width $2 r_0$ (here, $S_{D} = 2\pi^{D/2}$
$R^{(D-1)}/\Gamma(D/2) \sim R^{(D-1)}$ is the area of the
$D$-dimen\-sional sphere with radius $R$). Finally, we will
assume a particle density near to that of the liquid/crystal 
phase by fixing the mean particle separation $L = 
\rho^{-1/D}$ close to the interaction radius $R$, $L \sim R$.

Simple geometric considerations tell that in addition 
to the low-energy crystal configuration (characterized 
by space periodicity and a fixed number of nearest 
neighbours), the present model also develops numerous
metastable low-energy states with a disordered/glassy 
arrangement of particles. Such states are inhomogeneous 
in space and exhibit a notably smaller average number 
of nearest neighbours as compared to the crystal. Also, 
geometric considerations tell that such glassy 
configurations are `well separated' from the ordered 
crystal: their transformation into the crystal state 
would require a global (on the scale of the entire 
system) rearrangement of the particles which would 
involve large energies. It is this type of random 
glass-like configurations which is at the focus of 
our further studies below.

\section{Replica free energy}\label{sec:rfenergy}

We start from the expression (\ref{mfg22}) for the replica
partition function in the form
\begin{eqnarray}
   \label{mfg27a} 
   && Z_{m} = \frac{m^{DN} V^{N}}{N!}
   \prod_{i=1}^{N}
   \prod_{a=1}^{m} \int d^{D} {\bf u}_{i}^{a}
   \, \delta\Bigl(\sum_{b=1}^{m} {\bf u}_{i}^{b}\Bigr) \\
   && 
   \times\! \prod_{i=1}^N\! \int\! \frac{d^D {\bf x}_i}{V}
   \exp\Bigl\{-\frac{\beta}{2}\sum_{a=1}^{m} 
   \sum_{i,j=1}^N \!  U\bigl[|({\bf x}_i\!-\!{\bf x}_j)\!
   +\!{\bf u}_{ij}^a|\bigr]\!\Bigr\},
   \nonumber
\end{eqnarray}
where ${\bf u}_{ij}^a \equiv ({\bf u}_i^a-{\bf u}_j^a)$. 
For large $N$, we can approximate $N! \sim N^N$ and 
obtain the prefactor $m^{DN} V^N/N! \simeq \exp[ D N 
{\rm ln}(m) - N {\rm ln}(\rho)]$ with the density 
$\rho = N/V$. The replica partition function then 
takes the form 
\begin{eqnarray}
   \label{mfg27} 
   && Z_{m} = e^{\left[ D N {\rm ln}(m) - N {\rm ln}(\rho) \right]}
   \prod_{i=1}^{N}
   \prod_{a=1}^{m} \int d^{D} {\bf u}_{i}^{a}
   \, \delta\Bigl(\sum_{b=1}^{m} {\bf u}_{i}^{b}\Bigr) 
   \nonumber \\ 
   && \qquad\qquad\qquad\qquad \times 
   \bigl\langle \exp\left(-\beta H\left[{\bf x};{\bf u}\right]\right) 
   \bigr\rangle_{{\bf x}} 
\end{eqnarray}
with the average over positions
\begin{equation}
   \label{mfg27b} 
   \langle \Phi\left[{\bf x}_1, \dots ,
   {\bf x}_N\right] \rangle_{{\bf x}} \equiv
   \prod_{i=1}^{N}\int \frac{d^{D} {\bf x}_{i}}{V}
   \Phi\left[{\bf x}_1, \dots , {\bf x}_N\right].
\end{equation}
The average of the exponential in Eq.\ (\ref{mfg27}) can be calculated 
with the help of a cumulant expansion,
\begin{eqnarray}
   \label{mfg29} 
   &&\bigl\langle\exp\left(-\beta H \left[{\bf x}; {\bf u}\right] \right) 
   \bigr\rangle_{\bf x} 
   \\ && 
   \simeq \exp\Bigl[-\beta
   \langle H\rangle_{\bf x} + \frac{\beta^{2}}{2!} \beta^{2}
   \langle\langle H^{2}\rangle\rangle_{\bf x} -
   \frac{\beta^{3}}{3!}\langle\langle H^{3}\rangle\rangle_{\bf x} 
   + \; ... \; \Bigr],
   \nonumber
\end{eqnarray}
where $\langle\langle H^{k}\rangle\rangle_{\bf x}$ denotes the $k$-{\it th}
cumulant of the Hamiltonian; in particular,
\begin{eqnarray}
\label{mfg30}
  \langle\langle H^{2}\rangle\rangle_{\bf x} 
  &\equiv & 
  \langle H^{2}\rangle_{\bf x} - \langle H\rangle_{\bf x}^{2},
  \\
  \noalign{\vskip 3pt} 
  \langle\langle H^{3}\rangle\rangle_{\bf x} &\equiv & \langle
  H^{3}\rangle_{\bf x} - 3 \langle H^{2}\rangle_{\bf x}
  \langle H \rangle_{\bf x} + 2 \langle H\rangle_{\bf x}^{2}.
  \label{mfg30a}
\end{eqnarray}

\subsection{First- and second-order contributions}

\begin{figure}[h]
\includegraphics[scale=0.33]{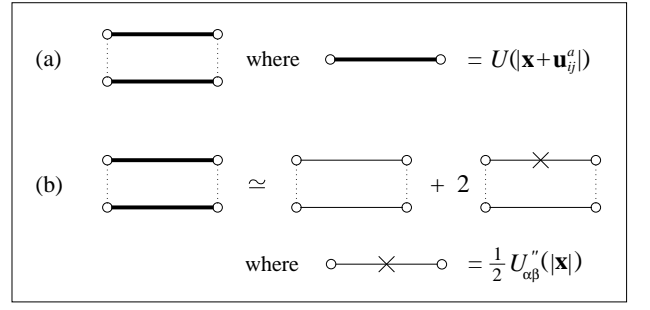} 
\caption[]{Diagrammatic representation of the second-order cumulant;
  (a) only connected graphs give a finite contribution,
  (b) contributions after expansion in the displacement 
   ${\bf u}_{ij}^a$. }
  \label{fig:sec_ord_dia}
\end{figure}

Using the potential (\ref{mfg24}) with its volume integral 
(\ref{mfg25}), we find the first-order term
\begin{eqnarray}
   \label{mfg32} 
   \langle H\rangle_{\bf x}  
   \!&=&\! \frac{1}{2}\sum_{a=1}^{m} \sum_{i,j=1}^{N}
      \langle U\left[|({\bf x}_{i} - {\bf x}_{j}) +
      {\bf u}_{ij}^{a}|\right]\rangle_{\bf x} \\
   \!&=&\! \frac{N^2}{2}m \int \frac{d^{D} {\bf x}}{V}
      U\left(|{\bf x}|\right) 
   \simeq - \frac{N\rho V_0}{2}m . \nonumber
\end{eqnarray}
For the second-order cumulant (see Fig.\ \ref{fig:sec_ord_dia}(a) 
for a diagrammatic representation), we find
\begin{eqnarray}
   \label{mfg33} 
   &&\langle\langle H^{2}\rangle\rangle_{\bf x}  
   = \frac{1}{4} \sum_{a,b=1}^{m} 
   \sum_{i,j,k,l}^{N} \langle\langle U[|({\bf x}_i-{\bf x}_j) 
   + {\bf u}_{ij}^a|] 
   \nonumber \\
   &&\qquad\qquad \times U[|({\bf x}_k-{\bf x}_l) 
   + {\bf u}_{kl}^b|]\rangle\rangle_{\bf x_i,x_i,x_j,x_k}.
\end{eqnarray}
Only correlated positions $(i,j) = (k,l)$ (i.e., connected 
graphs) give a finite contribution (since $ \langle 
U(|\delta {\bf x}_{ij} +{\bf u}_{ij}^b|) U(|\delta 
{\bf x}_{kl}+{\bf u}_{kl}^b|)\rangle =\langle 
U(|{\bf x}|)\rangle\langle U(|{\bf x}|)\rangle$), hence 
\begin{eqnarray}
   \label{mfg34} 
   \langle\langle H^2\rangle\rangle_{\bf x} 
   &=& \frac{1}{2} \sum_{a,b=1}^{m} \sum_{i,j}^{N} 
   \int \frac{d^{D} {\bf x}}{V} 
   \nonumber \\
   &&
   \times\, U\left(|{\bf x} + {\bf u}_{ij}^{a}|\right)\,
   U\left(|{\bf x} + {\bf u}_{ij}^{b}|\right).
\end{eqnarray}

We assume that typical deviations $\langle |{\bf u}|^2\rangle^{1/2}$
are small compared to $r_0$ (this assumption has to be verified at the
end of the calculation), allowing us to expand the potential 
$U[|({\bf x} + {\bf u}_{ij}^a|]$ in the displacement ${\bf u}_{ij}^{a}$,
\begin{eqnarray}
   \label{mfg36} 
   && \sum_{a=1}^{m} U(|({\bf x}+{\bf u}_{ij}^{a}|) 
   \simeq m U(|{\bf x} |) 
   + \sum_{\alpha=1}^{D} U_{\alpha}'(|{\bf x} |)
   \sum_{a=1}^{m} u_{ij (\alpha)}^a 
   \nonumber \\ &&\qquad\qquad
   + \frac{1}{2} \sum_{\alpha\beta=1}^{D} 
   U_{\alpha\beta}''(|{\bf x} |)
   \sum_{a=1}^{m} u_{ij (\alpha)}^a u_{ij (\beta)}^a.
\end{eqnarray}
Here, the indices $\alpha, \beta$ denote spatial vector 
components and $U_{\alpha}'$, $U_{\alpha\beta}''$ are 
corresponding derivatives.  The expansion (\ref{mfg36})
involves the small parameter  $u^{2}/r_0^2$, allowing 
its termination at the second order, cf.\ Fig.\ 
\ref{fig:sec_ord_dia}(b) for a diagrammatic representation.
Its second term (linear in $u$) vanishes due to the 
constraint (\ref{mfg21}) and we arrive at the expression 
\begin{eqnarray}
   \label{mfg37} 
   &&\langle\langle H^{2}\rangle\rangle_{\bf x} 
   \simeq \frac{N^2}{2} m^2 \langle 
   U^2(|{\bf x}|)\rangle_{\bf x} 
   \\ 
   && + \frac{1}{2}m \sum_{\alpha\beta=1}^{D} 
   \langle U(|{\bf x}|) U_{\alpha\beta}'' (|{\bf x}|)\rangle_{\bf x} 
   \sum_{a=1}^{m}\sum_{i,j,}^{N} u_{ij(\alpha)}^a u_{ij(\beta)}^a.
   \nonumber
\end{eqnarray}
Using the definition of the potential $U(|{\bf x}|)$, Eq.\ (\ref{mfg24}),
we obtain
\begin{equation}
   \label{mfg40} 
   U_{\alpha\beta}''( R_{\scriptscriptstyle -}\!<\!|{\bf x}|\!<\! 
   R_{\scriptscriptstyle +}) 
   \!=\!\frac{2}{r_0^2}
   \Bigl[\delta_{\alpha\beta}\Bigl(\!1-\frac{R}{|{\bf x}|}\Bigr) 
   + n_{\alpha}n_{\beta}\frac{R}{|{\bf x}|}\Bigr], 
\end{equation}
and 0 else, where $n_{\alpha} = x_{\alpha}/|{\bf x}|$ is the unit 
vector in the direction of ${\bf x}$. Accounting for the smallness
of $r_0$, $r_{0} \ll R$, we find the positional averages
\begin{eqnarray}
   \label{mfg38} 
   \langle U(|{\bf x}|)^2\rangle_{\bf x}
   &\equiv& \int \frac{d^{D} {\bf x}}{V} \bigl[U(|{\bf x}|)\bigr]^{2} 
   \simeq \frac{V_{0}}{V},
   \\
   \label{mfg41} 
   \langle U(|{\bf x}|) U_{\alpha\beta}''(|{\bf x}|)\rangle_{\bf x}  
   &\equiv& \int \frac{d^{D} {\bf x}}{V} U(|{\bf x}|)
   U_{\alpha\beta}''(|{\bf x}|) 
   \\
   &\simeq& 
   -\frac{V_{0}}{V} \frac{2}{r_{0}^{2}} \langle
   n_{\alpha}n_{\beta}\rangle = -\frac{V_{0}}{V}
   \frac{2}{r_0^2 D} \delta_{\alpha\beta}, \nonumber
\end{eqnarray}
and substituting these results back into Eq.\ (\ref{mfg37}) 
we obtain
\begin{eqnarray}
   \label{mfg42} 
   \langle\langle H^{2}\rangle\rangle_{\bf x} 
   &\simeq& \frac{N\rho V_0}{2} m^2 
   \\ &-& \frac{N\rho V_0}{D r_0^2}m  
   \sum_{i,j}^{N} \sum_{\alpha=1}^{D} 
   \sum_{a=1}^{m} \bigl[ u_{ij (\alpha)}^{a}\bigr]^{2}.
   \nonumber
\end{eqnarray}
Going back to the original displacement coordinates 
${\bf u}_{i}^{a}$ and accounting for the restrictions 
(the second condition corresponds to a global shift 
of all particles in the system) $\sum_{a=1}^{m} 
{\bf u}_{i}^{a} = 0$ and $\sum_{i}^{N} {\bf u}_{i}^{a} 
= 0$, we obtain the following contribution from the 
first- and second-order cumulants,
\begin{eqnarray}
   \label{mfg44} 
   &&-\beta \langle H\rangle_{\bf x} + 
   \frac{\beta^2}{2!} \langle\langle H^{2}\rangle\rangle_{\bf x}
   \\
   && \qquad\qquad\qquad  \simeq \frac{N\rho V_0}{2} 
   \Bigl[ (\beta m) + \frac{1}{2!} (\beta m)^{2} \Bigr] 
   \nonumber \\
   && \qquad\qquad\qquad 
   - \frac{\rho V_0 \beta}{D r_0^2} (\beta m) 
   \sum_{\alpha=1}^{D} \sum_{a=1}^{m}
   \sum_{i}^{N} \bigl[ u_{i (\alpha)}^{a}\bigr]^{2}.
   \nonumber
\end{eqnarray}

\subsection{Third-order cumulant}

\begin{figure}[h]
\includegraphics[scale=0.30]{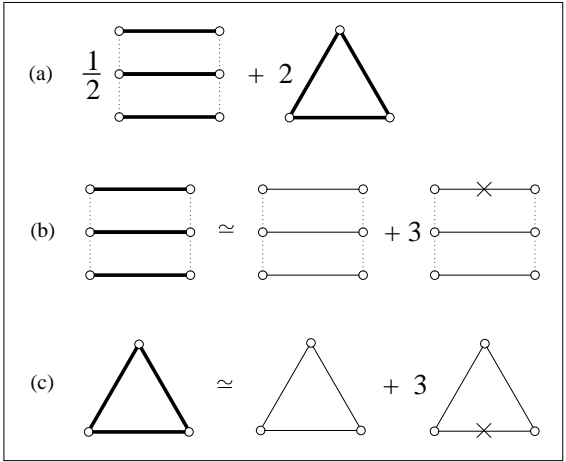}
\caption[]{Diagrams contributing to third order in the 
  interaction potential: 
  (a) two-point and three-point (loop) diagrams,
  (b) contribution of the two-point diagram after 
  expansion in the displacement ${\bf u}$,
  (c) contribution of the three-point diagram after
  expansion in the displacement ${\bf u}$.}
  \label{fig:thi_ord_dia}
\end{figure}

Next, we determine the contributions from the third-order
cumulant $\langle\langle H^{3}\rangle\rangle_{\bf x}
\equiv (\Xi/2 + 2\Delta)$ which contributes 
with two terms describing two-point ($\Xi$) and 
three-point ($\Delta$) correlations. These correspond to the 
connected diagrams shown in Fig.\ \ref{fig:thi_ord_dia}(a) 
and can be written in the form
\begin{equation}
   \label{mfg45} 
   \Xi =
   \sum_{a,b,c=1}^{m} \sum_{i,j}^{N} 
   \langle U[| {\bf x} + {\bf u}_{ij}^{a}|] 
   U[| {\bf x} + {\bf u}_{ij}^{b}|]
   U[| {\bf x} + {\bf u}_{ij}^{c}|]\rangle_{\bf x}
\end{equation}
and
\begin{eqnarray}
   \label{mfg46} 
   &&\Delta = \sum_{a,b,c=1}^{m}
   \sum_{i,j,k}^{N} \langle 
   U[|({\bf x}_{i} - {\bf x}_{j}) + {\bf u}_{ij}^{a}|] 
   \\
   && \quad \times
   U[|({\bf x}_{j} - {\bf x}_{k}) + {\bf u}_{jk}^{b}|] 
   U[|({\bf x}_{k} - {\bf x}_{i}) + {\bf u}_{ki}^{c}|]
   \rangle_{{\bf x}_{i}{\bf x}_{j}{\bf x}_{k}}.
   \nonumber
\end{eqnarray}
Expanding the two-point contribution Eq.\ (\ref{mfg45}) in 
the displacement ${\bf u}$ and integrating over ${\bf x}$, 
we obtain up to second order in the displacement ${\bf u}$ 
(cf.\ Fig.\ \ref{fig:thi_ord_dia}(b))
\begin{equation}
   \label{mfg48}
   \Xi = - N\rho V_0 m^3
   +\frac{6\rho V_0}{D r_0^2}m^2
   \sum_{\alpha=1}^{D} \sum_{a=1}^{m}
   \sum_{i}^{N} \bigl[u_{i (\alpha)}^{a}\bigr]^{2}\!\!.
\end{equation}
Next, we concentrate on the three-point (loop) contribution 
and first simplify the expression by redefining the coordinates 
${\bf x}_{i}, {\bf x}_{j}$, ${\bf x}_{k}$, 
\begin{eqnarray}
   \label{mfg47} 
   &&\Delta = \sum_{a,b,c=1}^{m}
   \sum_{i,j,k}^{N} \langle 
   U[|{\bf x}_{1} + {\bf u}_{ij}^{a}|]
   \\
   && \qquad\qquad \times
   U[|{\bf x}_{2} + {\bf u}_{jk}^{b}|]
   U[|{\bf x}_{1} - {\bf x}_{2} + {\bf u}_{ik}^{c}|]
   \rangle_{{\bf x}_{1} {\bf x}_{2}}.
   \nonumber
\end{eqnarray}
We expand in the displacement ${\bf u}$, cf.\ Fig.\
\ref{fig:thi_ord_dia}(c), and arrive at the expression
\begin{eqnarray}
   \label{mfg49} 
   &&\Delta = N^3 m^3 \langle
   U(|{\bf x}_1|) U(|{\bf x}_2|)
   U(|{\bf x}_1-{\bf x}_2|)\rangle_{{\bf x}_1 {\bf x}_2} 
   \\
   && +\frac{3N}{2}m^2 \sum_{\alpha\beta=1}^{D}
   \langle U(|{\bf x}_{1}|) U(|{\bf x}_{2}|)
   U_{\alpha\beta}''(|{\bf x}_1-{\bf x}_2|)
   \rangle_{{\bf x}_{1} {\bf x}_{2}} 
   \nonumber \\
   &&\qquad\qquad\qquad\qquad\qquad\qquad  \times
   \sum_{a=1}^{m} \sum_{ij}^{N} u_{ij (\alpha)}^{a} 
   u_{ij (\beta)}^{a}.
   \nonumber
\end{eqnarray}
Using the definition (\ref{mfg24}) of the potential and its
derivatives (\ref{mfg40}), we find for the above disorder averages
(again assuming that $r_0 \ll R$) 
\begin{eqnarray}
   \label{mfg50} 
   && \langle U(|{\bf x}_{1}|) U(|{\bf x}_{2}|) 
   U(|{\bf x}_{1}-{\bf x}_{2}|) \rangle_{{\bf x}_1,{\bf x}_2}
   \simeq -\frac{V_{0} V_{1}}{V^{2}},\\
   && \langle U(|{\bf x}_{1}|) U(|{\bf x}_{2}|) 
   U_{\alpha\beta}''(|{\bf x}_{1}-{\bf x}_{2}|) 
   \rangle_{{\bf x}_1,{\bf x}_2}
   \simeq \delta_{\alpha\beta} \frac{2}{r_{0}^{2} D} 
   \frac{V_{0} V_{1}}{V^{2}}, \nonumber
\end{eqnarray}
with $V_{1} \sim R^{D-2} r_{0}^{2}$ the intersection volume 
of two spherical shells with radius $R$ and width $r_0$; 
the three-point contribution then takes the form
\begin{eqnarray}
   \label{mfg53} 
   \Delta &=& - N \rho^2 V_0 V_1 m^3 
   \\ && \qquad\qquad
   + \frac{6 \rho^2 V_0 V_1}{D r_0^2} m^2
   \sum_{\alpha=1}^{D} \sum_{a=1}^{m}
   \sum_{i}^{N}  \bigl[ u_{i (\alpha)}^{a}\bigr]^{2}.
   \nonumber
\end{eqnarray}
Replacing the density $\rho$ with the distance parameter 
$L$ and assuming that $R \leq L$, we find that the
three-point contribution is reduced with respect to the 
two-point term (\ref{mfg48}) by the small factor 
\begin{equation}
   \label{mfg54} 
   \rho V_{1} \sim L^{-D} R^{D-2} r_{0}^{2} 
   = \left(\frac{R}{L}\right)^{D} 
   \left(\frac{r_{0}}{R}\right)^{2} \ll 1.
\end{equation}

\subsection{k-{\it th}-order cumulants}

\begin{figure}[h]
\includegraphics[scale=0.30]{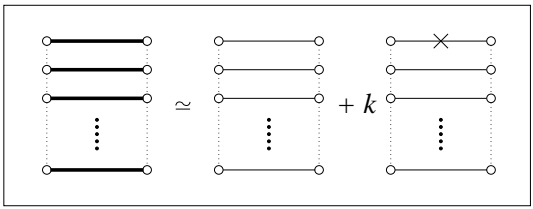}   
\caption[]{Diagrammatic representation of the two-point
  contribution to the $k$-{\it th}-order cumulant after 
  expansion in the displacement ${\bf u}$.}
  \label{fig:k_ord_dia}
\end{figure}

Comparing the magnitude of the above diagrams, we note that
all two-point terms (cf.\ Fig.\ \ref{fig:k_ord_dia}) appear 
with similar weight and we have to resum them; the two-point 
part of the $k$-{\it th}-order cumulant contributes with a 
term $(-\beta)^k \langle\langle H^k\rangle\rangle_{2\rm pt}/k!$ 
where ($k \geq 2$, see (\ref{mfg32}) for the $k=1$ contribution) 
\begin{eqnarray}
   \label{mfg56}
   (-1)^k\langle\langle H^{k}\rangle\rangle_{2\rm pt}
   \!&\simeq&\! 
   \frac{N \rho V_0}{2} m^k
   \\ \!&-&\! \frac{k\rho V_0}{D r_0^2}m^{k-1}
   \sum_{\alpha=1}^{D}
   \sum_{a=1}^{m} \sum_i^N\bigl[u_{i(\alpha)}^{a}\bigr]^{2}.
   \nonumber
\end{eqnarray}

On the other hand, the loop diagrams remain small: a
simple geometrical analysis shows that each $k$-{\it th}-order loop
involves an additional small factor of order ($k \geq 3$)
\begin{eqnarray}
   \label{mfg55} 
   (\rho V_{0})^{k-3} (\rho V_{1}) 
   &\sim& \biggl[\Bigl(\frac{R}{L}\Bigr)^{D} 
   \Bigl(\frac{r_{0}}{R}\Bigr)\biggr]^{k-3}
   \left(\frac{R}{L}\right)^{D} 
   \left(\frac{r_{0}}{R}\right)^{2}
   \nonumber \\ &=&
   \left(\frac{R}{L}\right)^{D(k-2)} 
   \left(\frac{r_{0}}{R}\right)^{k-1} \ll 1.
\end{eqnarray}
Thus, we conclude that for the present model with properly
chosen parameters $r_{0} \ll R$ and $R \leq L$ all loop-type 
contributions produce only small corrections as compared to the
main terms arising from the two-point diagrams.

It remains to sum up the two-point contributions, which corresponds
to the substitution $[\beta m + (\beta m)^2/2! + (\beta m)^3/3! \dots
\rightarrow [\exp(\beta m)-1]$, and we obtain the partition function
averaged over disorder in the form
\begin{eqnarray}
   \label{mfg57} 
   Z_{m}\!&=&\! \exp\bigl\{ D N {\rm ln}(m)\! 
   -\! N {\rm ln}(\rho) \!
   +\! (N\rho V_{0}/2) \bigl[e^{\beta m}\!-\!1\bigr]\bigr\} 
   \nonumber \\
   &\times&
   \prod_{i=1}^{N}\prod_{\alpha=1}^{D} 
   \biggl\{\prod_{a=1}^{m} \int d^{D} u_{i (\alpha)}^{a}\,
   \delta\Bigl(\sum_{b=1}^{m} u_{i (\alpha)}^{b}\Bigr)
   \\
   &\times&
   \exp\biggl[- \frac{\rho V_0 \beta}{D r_0^2} 
   \bigl[e^{\beta m} - 1\bigr]
   \sum_{a=1}^{m} \Bigl[u_{i(\alpha)}^{a}\Bigr]^2
   \biggr]\biggr\}.
   \nonumber
\end{eqnarray}
A simple Gaussian integration over the displacements 
$u_{i (\alpha)}^{a}$ then results in the expression
\begin{eqnarray}
   Z_{m} \!&=&\! \exp\biggl\{ \frac{DN}{2} {\rm ln}(m) -
    N {\rm ln}(\rho) 
   +\frac{N\rho V_0}{2}\bigl[e^{\beta m} - 1\bigr]
   \nonumber \\
   &-& \frac{N D (m-1)}{2} {\rm ln}
   \Bigl[\frac{\rho V_0 \beta}{D r_0^2} 
   \bigl[e^{\beta m} - 1\bigr] \Bigr] \biggr\}\label{mfg58}
\end{eqnarray}
and taking the logarithm we obtain (after rearranging 
terms) the replica free energy density $F(m,\beta) 
= -(1/N \beta m) {\rm ln}\left(Z_{m}\right)$ in the form, 
\begin{eqnarray}
   &&F(m,\beta) = F_{0}(\beta)
   -\frac{\rho V_{0}}{2 \beta m} \bigl[e^{\beta m} - 1\bigr] 
   \label{mfg59}\\ 
   && \quad
   -\frac{D}{2\beta m} \Bigl\{{\rm ln}\Bigl[\frac{\rho V_0 L^2 \beta m}
   {D r_0^2} \Bigr] 
   -(m-1) {\rm ln}\bigl[e^{\beta m}-1\bigr]\Bigr\},
   \nonumber
\end{eqnarray}
where we have separated the term 
\begin{equation}
   \label{mfg61a} 
   F_{0}(\beta) = \frac{D}{2\beta} {\rm ln}
   \left[\frac{\rho V_0\beta}{D r_0^2} \right]
\end{equation}
which does not depend on the replica number $m$. When proceeding
with (\ref{mfg59}) we have to keep in mind two restrictions
which are limiting its validity: {\it i)} our continuum
approximation prevents us from reaching very low temperatures,
and {\it ii)} the smallness of the mean square displacements  
$\langle u^2 \rangle \ll r_0^2$ requires a sufficiently
large parameter $\beta m$ or a sufficiently small temperature
$T$. We proceed with a detailed analysis of these restrictions.

\subsection{Restrictions}

In order to justify the continuum limit in the integrations over 
the displacements $u_{i (\alpha)}^{a}$, cf.\ Eq.\ (\ref{mfg57}), where
we assume that $|u_{i (\alpha)}^{a}| \gg 1$, we have to demand that 
the width 
\begin{equation}
   \label{mfg61c}
   \frac{\rho V_0\beta}{D r_0^2} 
   \bigl[e^{\beta m} - 1\bigr]\ll 1.
\end{equation}
Hence, the application of the continuuum limit puts a lower limit
on the system  temperature, or, alternatively, on the largeness of the
replica parameter $m$. We will return to this point in the next section.

The second restriction derives from the condition that typical 
values of the displacements $u_{i (\alpha)}^{a}$ as defined by the 
Boltzmann weight in Eq.\ (\ref{mfg57}) should be small on the scale
of the width $r_0$ of the attractive shell; the Gaussian
integration then yields the condition
\[
   \langle u^{2} \rangle \equiv 
   \frac{1}{m} \sum_{a=1}^{m} \bigl\langle[u^{a}]^{2}\bigr\rangle
   \sim \frac{D r_0^2/\rho V_0}
   {\beta m [\exp(\beta m) - 1]} \ll r_0^2.
\]
Since $\rho V_0 \sim (R/L)^D (r_0/R) \ll 1$ for $R \leq L$, 
we can conclude that the result for the replica free energy 
density describing the molecular liquid state is valid if 
$\beta m$ is sufficiently large and bounded by the condition
\begin{equation}
   \label{mfg69b} 
   \rho V_0\, \beta m \exp\left(\beta m\right) \gg 1.
\end{equation}
The violation of this condition implies that the particles, originally
assumed to be bounded in `replica molecules', would escape from 
the potential well $U(|{\bf x}|)$ and become effectively decoupled. 
In this situation the state of our replica system would not correspond
to the molecular liquid phase any more and the above analysis cannot
be applied. The same conclusion also follows from our  
general qualitative arguments in Sec.\ II.B: the limit of small
replica parameter $m$ corresponds to a high effective temperature 
of the replicated system where no `replica molecules' could survive.

\section{Results and conclusions}

\subsection{Free energy density and $T_c$}

Let us return to the free energy density (\ref{mfg59}) and 
determine the replica parameter $m$. Assuming that $\exp
\left(\beta m\right) \gg 1$, we first simplify the function 
$F(m,\beta)$, 
\begin{eqnarray}
   F(m,\beta)&\simeq& F_{0}(\beta) 
   -\frac{\rho V_0}{2 \beta m} 
   e^{\beta m}-\frac{D}{2\beta m} 
   {\rm ln} \biggl[\frac{\rho V_0 L^2 \beta m}{D r_0^2}\biggr] 
   \nonumber \\
   && \qquad\qquad\qquad +D (m-1)/2, \label{mfg61b}
\end{eqnarray}
and find its maximum in the replica parameter $m$ from the condition
$\partial_m F(m,\beta) = 0$,
\begin{equation}
   \label{mfg63a} 
   \frac{\rho V_{0}}{D}\left(\beta m - 1\right) e^{\beta m} 
   - \Bigl(\beta m^2 - 1\Bigr) =
   {\rm ln}\biggl[\frac{\rho V_0 L^2 \beta m}{D r_0^2}\biggr].
\end{equation}
With the parameters $r_0/L,\, \rho V_0 \ll 1$ and $\rho V_0 
L^2/r_0^2 \gg 1$, the above equation assumes the solution (to 
logarithmic accuracy) 
\begin{equation}
   \label{mfg64} 
   \beta m_{*}(\beta) \simeq {\rm ln}\frac{D}{\rho V_{0}} 
   \! + \! {\rm ln}
   \biggl[\frac{{\rm ln}(\rho V_0 L^2/e D r_0^2)}
   {{\rm ln}(D/\rho V_0)} 
   + \frac{1}{\beta} {\rm ln}\frac{D}{\rho V_0}\biggr].
\end{equation}
In the following, we concentrate on the case of matching
density $\rho \sim R^{-D}$, i.e., $L \sim R$, and 
substituting $\rho \sim R^{-D}$, $V_0 \sim R^{D-1} r_0$ 
we easily verify that the above conditions are satisfied, 
\begin{eqnarray}
\label{mfg65} 
   \rho V_{0} &\sim& \left(\frac{R}{L}\right)^{D}
   \frac{r_{0}}{R} \sim \frac{r_{0}}{R} \ll 1,
   \\
   \label{mfg65a} 
   \rho V_{0} \frac{L^{2}}{r_{0}^{2}} 
   &\sim& \left(\frac{R}{L}\right)^{D-1} \frac{L}{r_{0}} 
   \sim \frac{R}{r_{0}} \gg 1.
\end{eqnarray}
The solution $m_*$ then assumes the simplified form
\begin{equation}
   \label{mfg66} 
   \beta m_{*}(\beta) \sim {\rm ln}\frac{DR}{r_{0}}
   + {\rm ln}\Bigl( 1 + \frac{1}{\beta} 
   {\rm ln}\frac{DR}{r_{0}} \Bigr).
\end{equation}
The transition temperature $T_c$ is defined by the condition 
$m_{*}(\beta_c) = 1$ (see Sec.\ II.B) and we arrive at the 
estimate, 
\begin{equation}
   \label{mfg67} 
   T_c \sim \frac{1}{{\rm ln}(D R/r_0)} \ll  1.
\end{equation}
Next, we should verify that the above conditions 
(\ref{mfg61c}) and (\ref{mfg69b}) are satisfied; they can 
be cast into the form
\begin{equation}
   \label{mfg69c}
   1 \ll \frac{r_0}{R} \beta m\, e^{\beta m} \ll m D r_0^2;
\end{equation}
with 
\begin{equation}
   \label{mfg68} 
   \beta m > \beta m_*(T) \sim {\rm ln}(D R/r_0).
\end{equation}
The first relation (guaranteeing a small displacement amplitude 
$\langle u^2\rangle \ll r_0^2$) is satisfied within the entire 
low-temperature phase $0 \leq T \leq T_c$. The second relation
(allowing us to make use of the continuum limit) implies that
$T \gg 1/r_0^2$. This condition is satisfied at $T_c$ since
$\ln(DR/r_0) \ll r_0^2$ for sufficiently large $r_0$; however,
it is violated at low temperatures $T \sim 1/r_0^2$, thus 
limiting the applicability of our results to sufficiently
high values of $T$.

The logarithmic dependence of $T_c$ found above can be
easily understood from an order of magnitude estimate 
of the free energies of the liquid and glass phase. 
Let us assume that the average number of particles 
interacting in the frozen state is of the order of $D$ 
(see below, Eqs.\ (\ref{EG}) and (\ref{EGA})). Then each 
particle contributes with an energy $-D$ 
to the free energy of the system. Second, the freezing
into the glass state confines the particle to the volume
$r_0^D$ which is small with respect to the volume $R^D$
available in the liquid state. This produces a deficit
$\ln(R^{D}/r_{0}^{D})$ in the entropy of the frozen state
as compared to the liquid. The difference in the free
energies between the frozen and liquid states then can
be estimated as $\Delta F \sim -D + T \ln(R^{D}/r_{0}^{D})$;
this quantity turns negative at temperatures $T < T_{c}
\sim [\ln(R/r_{0})]^{-1}$ and the frozen state becomes
preferable.

An independent confirmation of the result (\ref{mfg67}) can be
obtained from a virial expansion \cite{LL}: to third order 
the equation of state takes the form
\begin{equation}
   \label{virial}
   \frac{pV}{NT} = 1+B(T)\frac{N}{V}+C(T)\frac{N^2}{V^2}+\dots
\end{equation}
with the coefficients 
\begin{eqnarray}
   \label{virialcoeff}
   B(T) &=& -\frac{1}{2} \int d^3 x \, f({\bf x}), \\
   C(T) &=& -\frac{1}{3} \int d^3 x_1 d^3 x_2 \, f({\bf x}_1) 
          f({\bf x}_2) f({\bf x}_1 - {\bf x}_2), \nonumber
\end{eqnarray}
and $f({\bf x}) = \exp[-\beta U({\bf x})] -1$. Inserting the
expression (\ref{mfg24}) for the potential $U$ we can
rewrite
\begin{equation}
   \label{virf}
   f({\bf x}) \simeq \left\{\begin{array}{ll}
   \exp(\beta), &
   R_{\scriptscriptstyle -}<|{\bf x}|<R_{\scriptscriptstyle +},\\
   \noalign{\vskip 3 pt}
   0, &
   |{\bf x}| \not\in 
   [R_{\scriptscriptstyle -},R_{\scriptscriptstyle +}],\\
   \end{array}
   \right.
\end{equation}
and using the results (\ref{mfg25}) and (\ref{mfg50}) above,
we find $B \simeq -R^D(r_0/R) \exp(\beta)$ and $C(T) \simeq 
-R^{2D}(r_0/R)^3 \exp(3\beta)$. Assuming a high particle density 
with $L\sim R$, we find that the virial expansion diverges
when $(r_0/R)\exp(\beta) \sim 1$, thus reproducing the critical
temperature (\ref{mfg67}).

\subsection{Configurational entropy}

We can use our free energy expression $F(m,\beta)$
to construct the form of the configurational entropy $S(f,T)$, 
cf.\ Fig.\ 1. Taking the derivative of the free energy 
$F(m,T) = f(m,T)-(T/m)S[f(m,T),T]$ and of $mF(m,T)$ with 
respect to $m$ and using that $\partial_f S = m/T$,
cf.\ (\ref{Fml}) and (\ref{dS}), we obtain the relations
\begin{eqnarray}
   \frac{m^2}{T} \partial_m F(m,T) = S(m,T), \label{confS} \\
   \partial_m [mF(m,T)] = f(m,T), \label{conff}
\end{eqnarray}
and eliminating the parameter $m$, we arrive at an expression 
for $S(f,T)$. We use the free energy (\ref{mfg59}) in the limit
of large $\beta m$ and expand (to third order in $m-m_*$) 
around the maximum $F(m_*(T),T)$. Expressing $m-m_*$ through
$f-f_{\mathrm{min}}$ (with $f_{\mathrm{min}} = F(m_*(T),T)$)
we find that the configurational entropy near $f_{\mathrm{min}}$
assumes the form
\begin{equation}
   S(f,T) \approx \beta m_*(f-f_{\mathrm{min}})
   -\frac{2R}{r_0 T} e^{-\beta m_*} (f-f_{\mathrm{min}})^2.
   \label{confSf}
\end{equation}
This result exhibits the shape expected for a configurational
entropy triggering an entropy crisis as discussed in Sec.\ II.B,
see Fig.\ \ref{fig:sf}.

\subsection{Order parameter}

The above formal replica computations have been performed along the 
lines outlined in section II and fit well the original physical 
ideas regarding the nature of the liquid-glass phase transition. 
In order to describe the phase transition in more quantitative 
terms one has to introduce a properly defined order parameter 
which should be a measureable quantity, e.g., in computer 
simulations. Such an order parameter is easily defined if 
we exploit the replication trick where we have introduced 
$m$ identical copies of the same system.  Let us assume that
all these systems are allowed to thermalize independently but
with the same (random) starting positions of the particles. 
If, at sufficiently high temperatures, the thermodynamic state 
of the system is a liquid, we expect the spatial positions of 
the particles belonging to different systems to be uncorrelated.
On the other hand, if the thermodynamic state of the system
is a frozen glass with each particle localized in a limited part 
of space, then the positions of the particles belonging 
to different replicas remain correlated, although the systems 
are uncoupled.  Keeping in mind this qualitative scenario, 
we introduce the correlator
\begin{equation}
   \label{mfg70} 
   G = \frac{1}{N} \sum_{i=1}^{N} 
   \bigl\langle\bigl( {\bf x}^{\scriptscriptstyle (1)}_i
   -{\bf x}^{\scriptscriptstyle (2)}_i\bigr)^2\bigr\rangle,
\end{equation}
where ${\bf x}^{\scriptscriptstyle (1)}_{i}$ is the 
position of the $i$-{\it th} particle in the system number 1, 
and ${\bf x}^{\scriptscriptstyle(2)}_{i}$ is the corresponding
position (again of the $i$-{\it th} particle) of the system number 2.
If the thermodynamic state of the system is a liquid, then the 
value of $G$ is proportional to $N^{-1} \sum_{i=1}^{N} 
\bigl\langle\bigl({\bf x}^{\scriptscriptstyle (1)}_{i}\bigr)^{2}
\bigr\rangle$, which is of the order of the linear size squared 
of the system and becomes infinite in the thermodynamic limit. 

On the other hand, if the system is in the glass state, the 
situation is very different. In this case both 
${\bf x}^{\scriptscriptstyle (1)}_{i}$ and 
${\bf x}^{\scriptscriptstyle (2)}_{i}$ are localised near 
{\it the same} equilibrium positions and $G$ remains
finite. Since all $m$ systems (replicas) considered here are 
equivalent, it is convenient to write the correlator 
(\ref{mfg70}) in the symmetric way
\[
   G = \frac{1}{N} \sum_{i=1}^{N}  
   \frac{1}{m(m-1)} \sum_{a,b=1}^{m} 
   \bigl\langle\bigl({\bf x}^{\scriptscriptstyle (a)}_i
   -{\bf x}^{\scriptscriptstyle (b)}_i\bigr)^{2}\bigr\rangle.
\]
We compute $G$ in the glass phase following the procedure 
described in section II and implemented in section III. 
Changing variables according to ${\bf x}_{i}^{a}={\bf x}_{i}
+{\bf u}_{i}^{a}$, we find
\begin{eqnarray}
   G &=& \frac{1}{m(m\!-\!1)} \! \sum_{a,b=1}^{m} \!
   \bigl\langle\bigl({\bf u}^{a} - {\bf u}^{b}\bigr)^{2}\bigr\rangle
   \nonumber \\
   &=& \frac{2 D m}{(m-1)}
   \bigl\langle\left(u^a_{\alpha}\right)^2\bigr\rangle,
   \label{mfg72} 
\end{eqnarray}
with the distribution of displacements $u^{a}_{\alpha}$ given
by Eq.\ (\ref{mfg57}). A simple Gaussian integration yields
\begin{equation}
\label{mfg73} 
   G \simeq T \frac{D r_0^2}{\rho V_0} e^{-\beta m_{*}(T)},
\end{equation}
where $\rho V_0 \sim r_0/R$ (see Eq.\ (\ref{mfg65})) and 
the value of $m_{*}(T)$ is given by the saddle-point 
solution $\beta m_{*}(T) \sim {\rm ln}\left(D R/r_0\right)$. 
Thus, in the glassy phase we find the finite value
\begin{equation}
   \label{mfg74} 
   G \simeq T  r_0^2.
\end{equation}
We then define the order parameter $Q = (1 + G)^{-1}$ assuming 
a finite value in the glassy phase and vanishing in the liquid,
\begin{equation}
   \label{mfg75}
   Q = \left\{\begin{array}{ll}
             0, &
             \mbox{in the liquid, at}  \;  T  >  T_{c},\\
             \noalign{\vskip 5 pt}
             \displaystyle{\frac{1}{1 + T r_{0}^{2}}},  &
             \mbox{in the glass, at}  \;   T  <  T_{c},
           \end{array}
   \right.
\end{equation}
where the transition temperature $T_{c}$ is given by Eq.\
(\ref{mfg67}). Note that the value of $Q$ remains finite at $T 
= T_c$, as expected for a phase transition driven by an entropy
crisis. On the one hand, the physical order parameter describing this 
phase transition exhibits a finite jump at $T_c$ (as expected for a 
first-order phase transition), while, on the other hand, the 
free energy of the system is continuous in the transition point 
(with a continuous derivative as in a second-order transition). 
On a qualitative level this is easily understood: in 
the present approach the glass phase is characterized by the 
typical size of spatial cells where the particles remain
localized. This volume is small in the low-temperature limit 
and grows with increasing temperature. At the transition to 
the liquid, the localization length is of the order of the 
interparticle distance and thus remains {\it finite}. Beyond 
the transition the particles move freely in the liquid
phase and the localization length is infinite. Hence the 
transition resembles the usual solid-liquid transition, 
however, with the jump in entropy (latent heat) replaced by
the entropy crisis scenario, guaranteeing the smooth transition
in the free energy density.

\subsection{Free energy and entropy densities}

Finally, we analyze in some more detail the value of the free 
energy- and entropy densities of the liquid and glassy phases.
The free energy density $F(m,\beta)$ of the liquid derives 
from the result (\ref{mfg59}) with the replica parameter $m=1$ 
(we assume that $\beta \gg 1$ and choose $L\sim R$),
\begin{equation}
   \label{mfg76}
   F_{\mathrm{liquid}}(\beta)  
   \simeq 
   -\frac{D}{2\beta} {\rm ln}(R^{2}) 
   -\frac{r_0}{2R \beta} e^\beta;
\end{equation}
the entropy density is given by the derivative
\begin{eqnarray}
   \label{mfg77}
   S_{\mathrm{liquid}}(\beta) &=&
   \beta^{2} \partial_\beta F_{\mathrm{liquid}}
   \\
   &\simeq&
   D {\rm ln}R 
   -\frac{r_0}{2R}(\beta-1)e^\beta.
   \nonumber
\end{eqnarray}
We observe that at sufficiently low temperatures the entropy density
of the liquid would become negative. Formally, this takes place at 
$\beta > \beta_{c}$ with $\beta_{c}$ defined by the condition
\begin{equation}
   \label{mfg78}
   (\beta_{c} - 1) e^{\beta_c} =   
   \frac{DR}{r_0} {\rm ln}(R^{2}).
\end{equation}
This negative entropy then signals the presence of the glass 
(or entropy crisis) transition: with $r_0/R \ll 1$ we obtain an
estimate for the transition temperature,
\begin{equation}
   \label{mfg79}
   \beta_c  \sim {\rm ln}\frac{D R}{r_{0}},
\end{equation}
which is in agreement with the analytic result (\ref{mfg67}).

The above arguments do not imply that the entropy of the liquid 
at the phase transition is equal to zero.  In fact, as we shall 
see below, the phase transition into the glassy phase takes place 
{\it before} the entropy becomes zero. The above consideration 
of the liquid entropy is merely a qualitative estimate of the  
temperature below which we would run into trouble if we were 
to use the result (\ref{mfg76}).

Turning to the frozen state, we first discuss the situation
deep in the glassy phase. The free energy density 
at $\beta \gg \beta_{c}$ follows from the result 
(\ref{mfg61b}) with $m = m_{*}(\beta)$ defined by 
(\ref{mfg66}). Assuming $L \sim R$ we find the simplified 
expression
\begin{eqnarray}
  F_{\mathrm{glass}} 
  &\simeq& -\frac{D}{2}\biggl[2 + \Bigl(1-\frac{1}{\beta}\Bigr)
   \frac{{\rm ln}{\rm ln}(DR/r_0)}{{\rm ln}(DR/r_0)} 
   \nonumber \\
   &&\qquad\qquad
   +\frac{1}{\beta}{\rm ln}\frac{r_0^2}{\beta}\biggr].
   \label{FG}
\end{eqnarray}
Taking the derivative with respect to temperature we obtain
the entropy density of the glassy phase, 
\begin{eqnarray}
   \label{mfg80}
   S_{\mathrm{glass}}(\beta) &=& 
   \beta^{2} \partial_\beta F_{\mathrm{glass}}
   \\ \nonumber
   &\simeq& \frac{D}{2} \left[
   1 + {\rm ln}\frac{r_{0}^{2}}{\beta}
   - \frac{{\rm ln}{\rm ln}(DR/r_0)}
   {{\rm ln}(DR/r_0)} \right].
\end{eqnarray}
Combining the results for the free energy and the entropy,
we can derive an expression for the average energy per particle 
in the glassy phase which scales with the dimension $D$,
\begin{equation}
\label{EG}
   E = F_{\mathrm{glass}} + T S_{\mathrm{glass}} 
   \simeq - A \, D;
\end{equation}
the prefactor $A$ assumes a value close to unity,
\begin{equation}
   \label{EGA}
   A \simeq 1 + \frac{{\rm ln}{\rm ln}(DR/r_0)}
   {2{\rm ln}(DR/r_0)} > 1.
\end{equation}
We thus find that our glass phase is `well packed' with 
slightly more than $D$ particles in a (interacting) 
nearest neighbor position on average. Note, that 
the entropy (\ref{mfg80}) becomes negative at sufficiently
low temperatures, i.e., when $\beta \sim r_{0}^{2} \gg 1$.
This unphysical result is a consequence of the breakdown 
of our continuum approximation at these low temperatures,
cf.\ Eq.\ (\ref{mfg61c}).

Next, we analyze the situation near the glass transition
temperature. A general expression for the entropy of the 
glassy phase (without assuming that $T \ll T_{c}$)
can be obtained from (\ref{mfg61b}),
\begin{eqnarray}
\label{mfg81}
  S_{\mathrm{glass}}(\beta) &\simeq& 
   - \frac{D}{2} {\rm ln}
  \frac{\beta}{DRr_0} 
  - \frac{D}{2} \beta m_{*}(\beta) \\
  \nonumber
  &\simeq& 
   D\,{\rm ln} R -\frac{r_0}{2R} 
   [\beta m_{*}(\beta) - 1] e^{\beta m_{*}(\beta)}
   \\
   &-& \frac{ D\beta m_{*}(\beta)}{2}  
   [1 - m_{*}]
   +\frac{D}{2} {\rm ln} m_{*}(\beta),
   \nonumber
\end{eqnarray}
where we have used the relation (\ref{mfg63a}) in order to
allow for a simple comparison with the result (\ref{mfg77}) 
of the liquid. At $T_c$ we have $m_{*} = 1$ and the 
entropy of the glass coincides with that of the liquid,
\begin{eqnarray}
   \label{mfg82}
   S_{\mathrm{glass}}(\beta_{c}) &=& S_{\mathrm{liquid}}(\beta_{c})
   = 
   \\
   &=& D\,{\rm ln}R 
   - \frac{r_0}{2R}\left(\beta_{c}-1\right)e^{\beta_c}
   \nonumber.
\end{eqnarray}
Inserting the estimate for $\beta_{c}$, Eq.\ (\ref{mfg67}), 
we find that the entropy is positive at the transition point,
\begin{equation}
   \label{mfg83}
   S_{\mathrm{glass}}(\beta_c) = S_{\mathrm{liquid}}(\beta_c) 
   \sim D \, {\rm ln} R - \frac{D}{2} {\rm ln} 
   \frac{R}{r_0} > 0.
\end{equation}

Another remark concerns the low-density or gas limit of
our model away from the densely packed limit with $L \sim R$
considered above. The system then does not exhibit any signature
of the above phase transition. Formally, we find no maximum
in the replica free energy density (\ref{mfg59}); rather,
$F(m,\beta)$ diverges for $m \rightarrow 0$. However, in this 
limit the constraint (\ref{mfg69b}) is violated and our 
calculation is not valid any more. In fact, the constraint
(\ref{mfg69b}) guarantees the smallness of the thermal
displacement amplitude $\langle u^2\rangle < r_0^2$; its
violation tells us that our basic assumption of a 
molecular liquid phase is corrupted, which signals that the
original system is in the liquid state. 

\subsection{Conclusion}

In conclusion, we have derived analytic results for the 
liquid-glass phase transition in a model glass former,
using the replica mean-field theory proposed by M\'ezard 
and Parisi \cite{MP} based on the entropy crisis scenario. 
Of course, the present study does not pretend to describe 
the physics in actual {\it realistic} glasses, which 
probably is much more complicated and furthermore 
essentially non-equilibrium in nature. Nevertheless, 
our analysis demonstrates that there exists a class of 
statistical systems which, on the one hand, incorporate 
essential features of structural glasses and, on the other 
hand, can be studied within an equilibrium statistical 
mechanics approach, similar to numerous other disordered 
systems. Although the present calculation was limited to 
a mean-field approximation, we have obtained a physically 
consistent set of results: Our replicated system indeed
exhibits a free energy $F(m,T)$ with a maximum 
$m_{*}(T)$ within the interval $[0,1]$ at low temperatures.
This maximum shifts towards unity with increasing $T$ and
the condition $m_{*}(T_c)=1$ defines a glass temperature 
$T_c \sim 1/{\rm ln}(DR/r_0)$ which is in agreement with 
various estimates. The maximum disappears upon diluting 
the system, indicating the persistence of the liquid state
for low densities $\rho \ll R^{-D}$. The behavior of the 
configurational entropy $S(f,T)$ is fully consistent 
with the freezing scenario based on an entropy crisis
and we recover the expected characteristics of a smooth
transition in the free energy combined with a jump in the
order parameter. In addition, our analysis provides a 
structural information on the glass state: calculating 
the mean particle energy from the free energy and entropy 
expressions, we find that the system freezes into a glass 
state with particles binding to slightly more than $D$ 
neighbors. All these results have been derived in analytic 
form, providing direct access to the parametric dependence 
of the physical results.

On the other hand, we have to admit that at the present 
stage it is difficult to juge which of the features of 
the phase transition scenario found here are specific to 
the particular model and its mean-field solution and 
which of them are generic and reflect the general 
situation encountered in the structural glass transition. 
At least part of this proviso can be handled by pushing
the analysis beyond the mean-field approximation.
With the experience gained in the present study, we 
believe that the theoretical construction of this next 
step, which would include the effect of thermodynamic 
fluctuations, does not look unrealistic.

We thank Lev Ioffe and Serguei Korshunov for interesting 
discussions and acknowledge financial support from the 
Swiss National Foundation through the Center of 
Theoretical Studies at ETH Z\"urich.

\end{document}